%% file: main.tex
\documentclass[journal=jctcce,manuscript=article,layout=traditional]{achemso}

\usepackage{graphicx}
\usepackage{helvet}
\usepackage[utf8]{inputenc}
\usepackage{hyperref}
\usepackage{xcolor}

\definecolor{tmavaseda}{gray}{0.2}
\hypersetup{
pdfborder={0 0 0 [0 0]},
colorlinks=true,
urlcolor=teal,
linkcolor=tmavaseda,
citecolor=tmavaseda
}

\title{Three stages of nascent protein translocation through the ribosome exit tunnel}

\author{Michal H. Kolář}
\affiliation{Department of Physical Chemistry, University of Chemistry and Technology, Technicka 5, 16628 Prague, Czech Republic}
\email{michal@mhko.science}

\author{Hugo McGrath$^\dagger$}
\affiliation{Department of Physical Chemistry, University of Chemistry and Technology, Technicka 5, 16628 Prague, Czech Republic}

\author{Felipe C. Nepomuceno$^\dagger$}
\affiliation{Department of Physical Chemistry, University of Chemistry and Technology, Technicka 5, 16628 Prague, Czech Republic}

\author{Michaela Černeková$^\dagger$}
\affiliation{Department of Physical Chemistry, University of Chemistry and Technology, Technicka 5, 16628 Prague, Czech Republic}

\begin{document}

\setlength{\parindent}{0em}
\setlength{\parskip}{0.7em}

\singlespacing

\maketitle

\begin{center}
\small
$^\dagger$ equal contribution
\end{center}

\begin{abstract}
All proteins in living organisms are produced in ribosomes that facilitate the translation of genetic information into a sequence of amino acid residues. During translation, the ribosome undergoes initiation, elongation, termination, and recycling. In fact, peptide bonds are formed only during the elongation phase, which comprises periodic association of transfer RNAs and multiple auxiliary proteins with the ribosome and the addition of an amino acid to the nascent polypeptide one at a time. The protein spends a considerable amount of time attached to the ribosome. Here, we conceptually divide this portion of the protein lifetime into three stages. We define each stage on the basis of the position of the N-terminus of the nascent polypeptide within the ribosome exit tunnel and the context of the catalytic center. We argue that nascent polypeptides experience a variety of forces that determine how they translocate through the tunnel and interact with the tunnel walls. We review current knowledge about nascent polypeptide translocation and identify several white spots in our understanding of the birth of proteins. 
\end{abstract}

\singlespacing

\section*{Introduction}

Ribosomes are the cornerstones of life as we know it because they are at the center of protein synthesis. Ribosomes catalyze the formation of peptide bonds and serve as hubs for fine-tuned regulation of the complex network of biochemical reactions enabled by ribosome-associated proteins. The ribosome is a ribonucleoprotein complex; it consist of ribosomal RNA (rRNA) and ribosomal proteins (r-proteins). The specific amount of these components varies among different organisms; in more advanced organisms, the proportion of r-proteins to rRNA is usually higher \cite{Melnikov12}. The ribosome is organized into two subunits: the large subunit contains the catalytic center, formed solely by rRNA, whereas the small subunit allows for the decoding of the messenger RNA (mRNA). Delivered by aminoacylated transfer RNAs (aa-tRNAs), the amino acids are connected one at a time deep in the ribosome.

Although the two main functions of the ribosome -- catalysis and decoding -- are highly evolutionarily conserved in the three domains of life, it is well established that the details of proteosynthesis in ribosomes are different between taxons \cite{Melnikov12}. The famous demonstration is the action of ribosome-targeting antibiotics that bind to bacterial ribosomes and modulate their action towards bacterium death, while keeping eukaryotic ribosomes functioning \cite{Wilson14}. Furthermore, ribosomes also differ within a single cell. Cytosolic ribosomes in eukaryotic cells are versatile and synthesize most of the cellular proteome. In contrast, mitochondrial ribosomes within the same cell are highly optimized to synthesize a small set of mitochondrial proteins \cite{Greber16}. In addition to this, ribosome populations appear heterogeneous in composition \cite{Norris21} and their ability to translate mRNAs \cite{Shi17}, leading to specialized ribosomes important, for example, in early embryogenesis \cite{Locati17}.

Translation takes place in four phases known as initiation, elongation, termination, and recycling \cite{Green97,Ramakrishnan02}. Each phase requires its own set of remarkably coordinated supporting proteins -- translation factors. Our understanding of the phases is based on a solid structural characterization of the ribosome and the translation-associated factors achieved over the last two decades that gained momentum after the first atom-resolved ribosome models were obtained \cite{Ban00,Schluenzen00,Wimberly00}.

During the initiation phase, the ribosome assembles with mRNA and the first tRNA, resulting in a complex where the tRNA that contains formylmethionine (fMet) in bacteria or methionine (Met) in archaea and eukaryotes is located on the start codon. This initiation complex enters the elongation phase, where aa-tRNAs are accommodated to the ribosome. The tRNAs sequentially translocate through three tRNA-binding sites spanning both subunits: aminoacyl (A), peptidyl (P), and exit (E) sites. Peptide bonds are formed between amino acids at the C-terminus of the nascent polypeptide (NP). Therefore, the N-terminus exits the ribosome first, while the C-terminus remains attached to the ribosome through the P-site tRNA. The release of NP is triggered by a stop codon that terminates the coding part of the mRNA. Stop codons are recognized by release factors capable of hydrolyzing an ester bond between tRNA and NP. The ribosome is made available again for other mRNA during the recycling stage, which involves dissociation of the ribosomal subunits.

\begin{figure}[tb]
    \centering
    \includegraphics{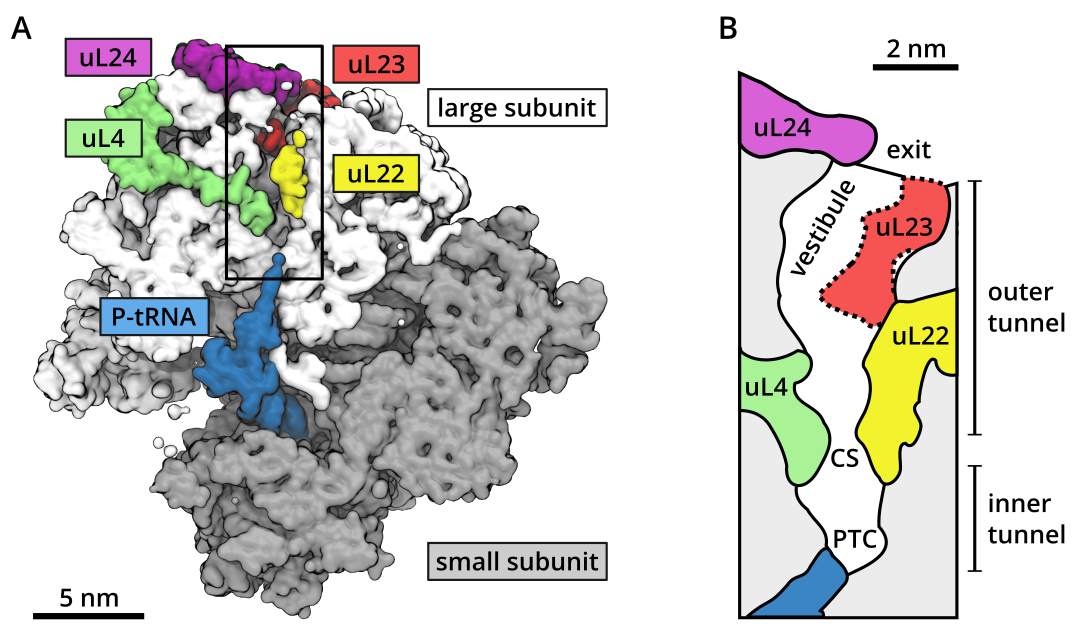}
    \caption{A) A cross-section of the bacterial ribosome highlighting the peptidyl tRNA (blue) and four ribosomal proteins lining the exit tunnel, uL4 (green), uL22 (yellow), uL23 (red), and uL24 (magenta). B) A diagram of the ribosome exit tunnel with the peptidyl-transferase center (PTC), constriction site (CS), vestibule and the exit port. In the literature, the inner and outer parts of the tunnel are sometimes called \emph{upper} and \emph{lower}, respectively, referring to the reverse tunnel orientation with the exit at the bottom.}
    \label{fig:overview}
\end{figure}

As proteosynthesis progresses, NP translocates through a void in the large ribosomal subunit known as the exit tunnel (Fig.\,\ref{fig:overview}). The translocation of the NP through the exit tunnel is regulated and has many physiological consequences. Protein folding is initiated within the tunnel \cite{Thommen17}. Furthermore, the tunnel binds to low-molecular-weight compounds that interfere with NP and can lead to translational arrest \cite{Ito13,Wilson16}. 

The tunnel is approximately 10\,nm long with a diameter between 0.8 and 1.5\,nm; however, the exact shape differs between organisms \cite{DaoDuc19}. Bacterial tunnels are generally longer and wider than tunnels in the ribosomes of higher organisms. The last third of the tunnel is wider and forms a vestibule (Fig.\,\ref{fig:overview}B). Although the inner tunnel is quite similar in all domains of life, the outer parts structurally diverge \cite{Gruschke10}. 

The structure, dynamics, and physicochemical properties of the exit tunnel determine how NPs leave the ribosome. Tunnel walls are primarily composed of rRNA, but also of a few r-proteins. In particular, in the first third of the tunnel, there is a constriction site (CS) formed by the r-proteins uL4 and uL22 that expose several evolutionarily conserved basic residues to the tunnel lumen \cite{Worthan22}. The vestibule is lined with the r-proteins uL23 and uL24 (Fig.\ref{fig:overview}). An additional r-protein eL39 is present in the vestibule of eukaryotic ribosomes. Furthermore, uL22 may have an extra loop that forms a second constriction in eukaryotes \cite{DaoDuc19}. Mitochondrial exit tunnels are still different. The ratio of r-proteins to rRNA in the mitoribosome is higher and the homologs of the exit tunnel r-proteins are generally much bulkier \cite{Sharma03a}. 

\begin{figure}
    \centering
    \includegraphics{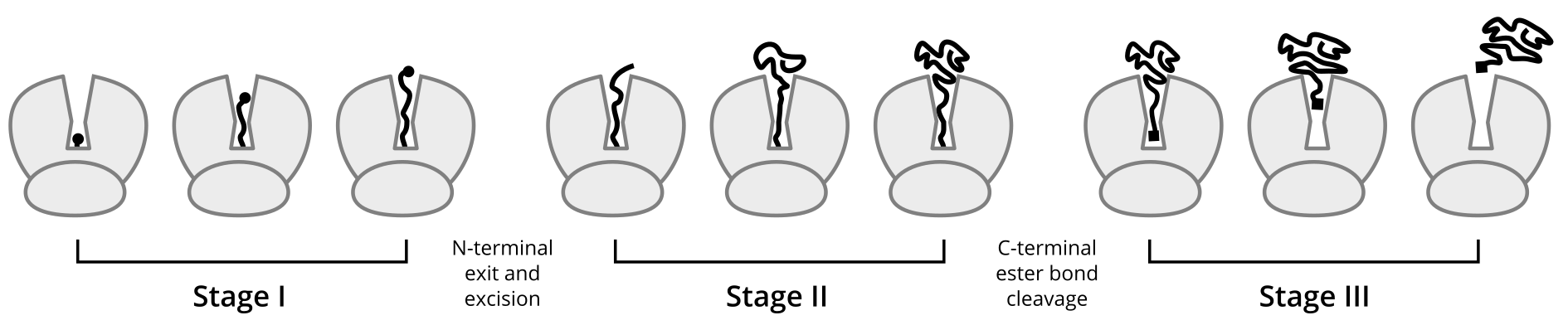}
    \caption{A diagram of three stages of the NP translocation through the exit tunnel. The black circle stands for the N-terminal fMet or Met that is cleaved off in the end of Stage I. The black square represents the C-terminus of NP released in the beginning of Stage III.}
    \label{fig:stages}
\end{figure}

The translocation of NP can be divided into three stages according to the position of the N-terminus of the NP and the context of the PTC (Fig.\ref{fig:stages}). We argue that the principles that govern the NP translocation differ in each stage. It seems that the three stages of translation are general for all three domains of life.

As described in detail in the following sections, during Stage I, the N-terminus of the NP travels from the PTC through the tunnel, presumably in the extended conformation. It passes through the CS and enters the vestibule of the outer part of the tunnel. After the N-terminal fMet or Met is cleaved off from the NP by ribosome-associated protein biogenesis factors, Stage II begins. It covers the time required to synthesize the remaining portion of the NP. The translocation of NP during Stage II is mainly driven by the folding of NP in the native tertiary structure outside the ribosome \cite{Cassaignau20}. Finally, Stage III represents the escape of the released C-terminus from the PTC to the surface of the ribosome.

Protein sequences vary considerably in length \cite{Zhang00}. In bacteria, the typical protein contains approximately 330 residues; in eukaryotes, it is 450, but the length distributions range from a few dozen to thousands or even tens of thousands of residues \cite{Reva08}. The translation rate depends on many factors, including the usage of synonymous codons, the structure of the mRNA, or NP--tunnel interactions \cite{Komar09}. With an average translation rate of 10--20 residues per second in bacteria and 4--5 in eukaryotes, protein synthesis can take less than a minute for small single-domain proteins to more than an hour for giant proteins. Consequently, the duration of the stages may vary, with the largest variations occurring in Stage II.

\section*{Stage I: Reaching the tunnel exit}

Stage I represents the time during which the N-terminus of the NP translocates from the PTC to the tunnel exit until it is chemically modified by enzymes known as ribosome-associated protein biogenesis factors. In bacteria, the first such enzyme is peptide deformylase (PDF), which removes the formyl group from the N-terminal fMet \cite{Adams68}. PDF binds to the surface of the ribosome near the tunnel exit and processes NP after about 50 residues have been synthesized \cite{Bingel-Erlenmeyer08}. However, the optimal length of an NP for PDF appears to be approximately 70 residues, as suggested by kinetic measurements and mathematical modeling \cite{Yang19}. In eukaryotes, only organellar PDFs have been identified \cite{Giglione01a}. The presence of PDFs in archaea remains a topic of debate \cite{Giglione01}. Archaea do not use fMet as the first residue, but they still contain an analog of PDF with unknown function \cite{Giglione04}.

Typically, bacterial deformylation is followed by the removal of the N-terminal Met, carried out by methionine aminopeptidase (MAP) \cite{Solbiati99}. Bacteria have one type of MAP, which can potentially bind to two sites on the ribosome near the tunnel exit, as indicated by a study of the \emph{E.\,coli} ribosome \cite{Bhakta19}. Eukaryotic cells comprise two types of MAP with N-terminal extensions, processing NP in both the cytoplasm and organelles \cite{Giglione04}. The evolutionary conservation of N-terminal excision is highlighted by the presence of MAP in archaea \cite{Tahirov98}. Similarly to PDFs, MAPs can act on NPs of about 50 residues long \cite{Koffer-Gutmann73,Yang19}.

The exit tunnel can protect about 35 NP residues from proteolytic enzymes \cite{Malkin67}. However, the tunnel plays an active role in the formation of compact structures \cite{Lu05a,Kolar22} so that up to 50 residues can be accommodated inside the ribosome \cite{Holtkamp15,Nilsson15}. $\alpha$-helices can already occur in the PTC \cite{Matheisl15,Su17a} but, as of now, this has only been shown for longer NPs. It is not clear whether $\alpha$-helices can form on short NPs during Stage I before the N-terminus leaves the tunnel. Depending on the level of NP compaction and the given average translation rates, we estimate that Stage I takes between 2 and 12 seconds. 

The main obstacle to the progress of NP through the tunnel is the CS, which is 0.8\,nm in diameter. Before the NP reaches the CS, no less than 10 residues are synthesized. The constriction represents an entropic barrier \cite{Yu23}; it remains a question of how the NP passes through it during Stage I.

The progression of short NPs through the tunnel and the translocation through the CS are facilitated by electrostatic interactions between the tunnel and NP \cite{Duc18}. This is in line with the fact that the N-termini of NPs are enriched by positively charged residues \cite{Tuller11}. However, the effect of positive NP residues on the translation rate is not universal. In later stages of elongation, positively charged portions of longer NPs often slow or even pause translation \cite{Lu08, Charneski13}. The pausing is supported by the secondary structure of the polyadenylate stretches of mRNA that is difficult to process by the ribosome \cite{Arthur15,Chandrasekaran19}. 

Beyond the CS, it seems that the wider tunnel allows for the translocation of NPs as a result of the increase in entropy. Duc and Song showed by the analysis of ribosome profiling data that the motion of NP can be described as a one-dimensional diffusion driven by electrostatics and entropy \cite{Duc18}.

\begin{figure}
    \centering
    \includegraphics{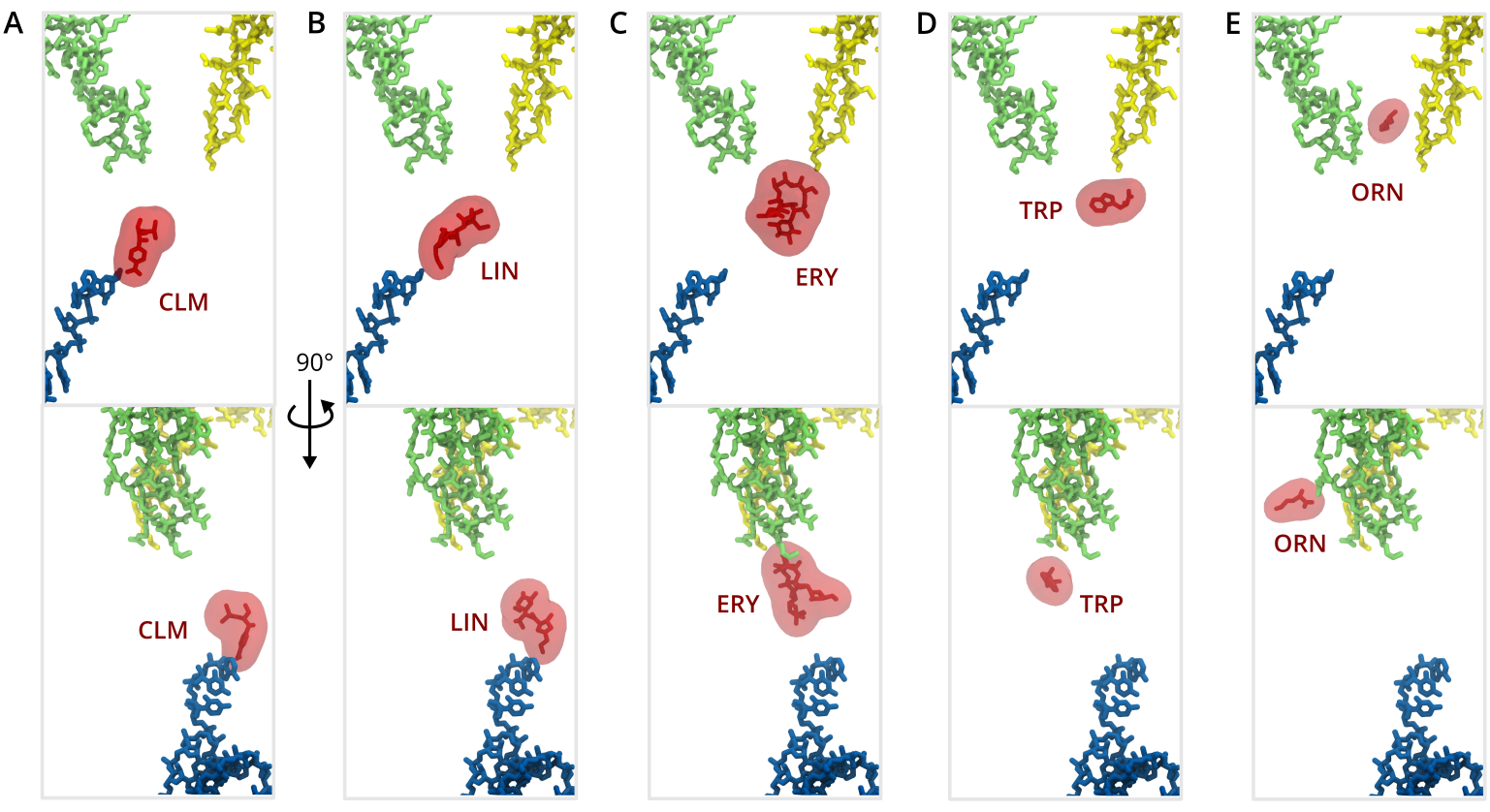}
    \caption{Binding sites of selected low-molecular-weight compounds (highlighted in red) within the ribosome exit tunnel. Ribosomal proteins uL4 (green) and uL22 (yellow), along with the P-site tRNA (blue), are shown for context. Antibiotics chloramphenicol (CLM) and lincomycin (LIN) occupy the A site, erythromycin (ERY) binds between the peptidyl transferase center and constriction site, while metabolites tryptophan (TRP) and ornithine (ORN) bind near the constriction site. Frontal and lateral views are presented in the upper and lower panels, respectively.}
    \label{fig:atbs}
\end{figure}

Short NPs can also interact with low-molecular-weight ligands during Stage I (Fig.\,\ref{fig:atbs}). Some bacterial genera employ ligand-dependent translational arrest to induce antibiotic resistance. For example, macrolides, such as erythromycin, can trigger translational arrest through specific interactions of the exit tunnel with NPs in a sequence-dependent manner \cite{Davis14, Beckert21}. This activates the expression of downstream genes such as \emph{ermC}, \emph{ermB}, or \emph{ermA}, which encode methyltransferases \cite{Koch17, Arenz16, Ramu11}. They add a methyl group to some 23S rRNA residues located in the ribosome tunnel and thus reduce the binding affinity of macrolides to the ribosomes (reviewed in Ref.\,\citenum{Weisblum95}). Other leader peptides arrest ribosomes in the presence of chloramphenicol, an antibiotic that binds to PTC \cite{Syroegin22}. This activates downstream genes encoding chloramphenicol acetyltransferases such as \emph{cmlA} \cite{Bissonnette91,Lovett96}, contributing to antibiotic resistance. 

Short NPs extend the functionality of the ribosome beyond catalysis to sense concentrations of some metabolites. For example, at higher concentration levels of L-tryptophan, the leader peptide TnaC in the exit tunnel helps ribosomes detect L-Trp (Fig.\,\ref{fig:atbs}D), leading to translation arrest at the termination point \cite{VanDerStel21,Su21}. SpeFL NP controls the expression of inducible ornithine decarboxylase in bacteria by helping the ribosome detect ornithine and arrest translation \cite{HerreroDelValle20} (Fig.\,\ref{fig:atbs}E). Viruses infecting eukaryotic cells utilize translation arrest to regulate their gene expression there. In human cytomegalovirus, the NP sequence of the leader peptide (22 residues) of UL4 gene regulates downstream translation of the glycoprotein gpUL4, with arrest promoted by NP interactions and eukaryotic release factor 1 (eRF1) during termination \cite{Janzen02}. 

Furthermore, during Stage I, tRNAs with short NPs may dissociate from the P site of the ribosome \cite{Menninger76}. The circumstances of this so-called peptidyl-tRNA drop-off are not fully understood, but it is assumed that it occurs due to weak interactions of NP with tunnel walls \cite{Das06}. Additionally, noncognate codon--anticodon pairing may contribute to the enhanced rate of tRNA dissociation. Based on a highly sensitive method for profiling peptidyl-tRNAs, it was suggested that drop-off is an active mechanism of protein synthesis quality control \cite{Nagao23}.

\section*{Stage II: On the ribosome}

We define Stage II as the second part of the elongation phase that starts after the N-terminus leaves the tunnel and is chemically modified by PDF and MAP. This usually happens after 50--70 residues are interconnected. The onset of Stage II is difficult to determine more precisely for two reasons. First, the length of the NP that leaves the tunnel depends on the NP sequence and factors that affect the cotranslational folding inside the tunnel. Second, N-terminal treatment by PDF and MAP is not universal; only about 55\% to 70\% of NPs are treated \cite{Giglione04}. Furthermore, some NPs undergo additional cotranslational chemical modifications of their N-termini, such as acetylation in eukaryotes \cite{Ree18}.

In Stage II, the NP folds into a tertiary structure. The proportion of proteins that fold this way is not insignificant; approximately a third of \emph{E.\,coli} cytosolic proteins exhibit cotranslational folding \cite{Ciryam13}. Some domains may already be folded within the vestibule without N-terminus treatment \cite{Nilsson15, Tu14, Kosolapov09}. Using fluorescence measurements, for example, it was shown that the tertiary structure of the nascent N-terminal domain of HemK in the vestibule differs from the native structure \cite{Holtkamp15}. Therefore, a refolding may occur outside the tunnel. On the other hand, some proteins only begin to fold once outside the exit tunnel \cite{Eichmann10}.

As the size of a protein grows, so does the probability of misfolding; multidomain proteins often misfold in solution \cite{Jahn16}. On the surface of ribosomes, one way in which the folding of multidomain proteins is biased towards the correct state is by the prevalence of rare slow translation codons at domain boundaries \cite{Komar09}. This allows more time for each domain to fold. N-terminal regions of the nascent multidomain proteins also interact with the ribosome surface, which guides them to the correct fold \cite{Liu17, Chan22}. 

In a crowded cellular environment, multisubunit protein complexes need to be assembled in a particular way to prevent aggregation or misfolding. In bacteria, genes that encode individual subunits are organized as operons in polycistronic mRNA, often in the order in which they are assembled \cite{Wells16}. This greatly increases the chances of successful arrangement of the subunits. Furthermore, mRNAs producing subunits that are to be assembled are often co-localized in the cell, increasing the local concentration of the relevant subunits \cite{Williams18}. In eukaryotes, translation, folding, and assembly are also integrated, as shown by the ribosome profiling of 12 protein complexes from \emph{S. cerevisiae} \cite{Shiber18}. Disome selective profiling in human cells also revealed that the formation of protein homo-oligomers occurs due to cotranslational assembly on adjacent ribosomes \cite{Bertolini21}.

The conformational energy landscape of NP is shaped not only by the ribosome itself \cite{Samatova24} but also by ribosome-associated chaperones \cite{Deuerling19}. Bacteria possess a chaperone known as trigger factor \cite{Hoffmann10}, higher organisms employ a complex network of chaperones that act on and off the ribosome \cite{Balchin16}. 

In any case, cotranslational protein folding generates a pulling force that drives the progress of NP through the tunnel. The force amounts to approximately 10 pN, as determined by optical-tweezers experiments using a SecM arrest peptide coupled with a \textit{de novo} designed protein \cite{Goldman15}. The pulling force is reduced when the trigger factor or GroEL chaperone is present, but only for proteins that fold on the surface of the ribosome, not in the vestibule \cite{Nilsson16}. In addition to cotranslational folding, a force is also generated during integration into and translocation through a membrane \cite{Cymer14, Ismail12}.

Proteins intended for integration into the bacterial plasma membrane or secretion have an N-terminal hydrophobic signal sequence, recognized by the ribonucleoprotein signal recognition particle (SRP) \cite{Luirink92} once 50--70 amino acids have been synthesized \cite{Kurzchalia86}. SRP interacts with the r-protein uL23 \cite{Gu03}. This system is highly conserved in all domains of life, used by eukaryotes to target NPs to the endoplasmic reticulum \cite{Walter82}, and archaea to the archaeal plasma membrane \cite{Bhuiyan00}.

In Stage II, the translocation of NP through the tunnel can be slowed down or even completely stopped. Translational arrest occurs due to a specific NP sequence of ribosome arresting peptides (RAP) \cite{Ito13}, or arrest is mediated by external ligands such as antibiotics or metabolites. The arresting NP sequences typically comprise up to 20 residues, not all of which are essential for the arrest. Single-molecule fluorescence experiments suggested that the stalling is tightly coupled to the dynamics of elongation \cite{Tsai14}.

Bacteria use RAPs as modulators of peptide translocation through membranes \cite{Nakatogawa02a, Ishii15}. These RAPs respond to the pulling forces that act on their emerging N-terminus. One of the well-studied examples is the SecM peptide that interacts with CS \cite{Bhushan11} and at the A site stabilizes aa-tRNA in the conformation that prevents peptide bond formation with NP \cite{Gersteuer24}. MifM-mediated arrest involves interactions of NP with CS and the region near the PTC, along with the N-terminus that interacts with outer ribosome parts \cite{Fujiwara18}. Vibrio export-monitoring polypeptide can adopt a highly compact conformation within the exit tunnel, avoiding the accommodation of tRNA at the A site \cite{Su17a}. In eukaryotes, X-box binding protein 1 (XBP1), involved in the unfolded protein response, can adopt a conformation within the exit tunnel to block PTC activity, causing temporary translation arrest \cite{Shanmuganathan19}. 

Eukaryotes possess RAPs that depend on small-molecule effectors that self-control their biosynthesis in Stage II. One such metabolite is S-adenosyl-L-methionine (AdoMet), which is detected by NP of an enzyme involved in the biosynthesis of methionine. A PEGylation assay suggested that the interaction of AdoMet with NP leads to a compact NP conformation and translational arrest \cite{Onoue11}. 

% intrinsic ribosome destabilization
Recently, NPs containing clusters of aspartate or glutamate residues have been associated with premature translation termination, a phenomenon termed intrinsic ribosome destabilization \cite{Chadani17}. It has been proposed that polyacidic NP stretches can cause the bacterial ribosome to assume a configuration that leads to noncanonical termination, although there is no direct structural characterization of this phenomenon yet. This type of termination occurs without the inclusion of release factors or ribosome splitting \cite{Chadani23}.

\section*{Stage III: Escape}

Once a stop codon is placed at the A site of the ribosome and a release factor recognizes it, a hydrolysis of the peptidyl-tRNA occurs, allowing the C-terminus of NP to move toward the tunnel exit \cite{Youngman08}. Hence, Stage III starts, the final part of the interaction between the NP and the ribosome. During Stage III, the ribosome tunnel ejects the C-terminal portion of NP, which is about 30 residues, according to the tunnel length \cite{Voss06}. This process has been extensively studied using computer simulations and theoretical models. Experimental results that would clarify the ejection of NP are very sparse.

The emergence of proteins released from the ribosome is driven by an entropic force generated by the motion of NP on the surface of the ribosome. This force is generated by the cotranslational folding of NP on the ribosomal surface \cite{Goldman15}. In addition, unstructured NP outside the tunnel may also generate a force, as determined by a PEGylation accessibility assay \cite{Fritch18}. Multiscale computer modeling in the same study further proposed that the pulling force is propagated to the C-terminus through the NP backbone \cite{Fritch18}.

Bui and Hoang have published a series of papers on this topic \cite{Bui16,Bui18,Bui20,Bui21,Bui23}. They have developed theoretical models and performed MD simulations at the coarse-grained level using tunnels of various complexity. They found that the escape time lies between microseconds and milliseconds, which is consistent with the results of other computer simulations of 122 \emph{E.\,coli} proteins \cite{Nissley20}. However, the experimental validation of the process timescale has yet to be completed.

The escape of NP can be quantitatively described by a simple diffusion model \cite{Bui16,Bui18,Bui20}. However, interactions between the tunnel and the NP appear to influence the ejection times, with the negatively charged C-terminus exiting more rapidly and positive charges decelerating the process \cite{Nissley20, Bui21}. This agrees with the ribosome profiling data from \emph{E.\,coli} \cite{Mohammad19} analyzed in Ref. \citenum{Nissley20}. They revealed that the presence of slowly ejecting sequences correlates with the time the ribosome spends on the stop codon. Hence, the NP escape may delay the onset of ribosome recycling \cite{Nissley20}.

Different organisms possess different geometries of ribosome exit tunnels, based on cryogenic electron microscopy (cryo-EM) reconstructions and X-ray crystallography structures of ribosomes \cite{DaoDuc19}. Therefore, escape times, even following the same diffusion mechanism \cite{Bui23}, may vary.

Short NPs cannot escape the tunnel, especially in the eukaryotes and archaea tunnels, because of the second CS present in these tunnels. Entrapment depends on the NP length; 12-mers, 20-mers, and 29-mers have a significantly higher chance of being trapped than other NP lengths, as observed in MD simulations in cylindrical models with varying radius \cite{Yu23}. Similar results were obtained by simulations using a ribosomal exit tunnel from \textit{H. marismortui} \cite{Bui20}. Simulation experiments using coarse-grained exit tunnels from the three domains of life \cite{Chwastyk21} as well as simpler exit tunnel models \cite{Yu23} showed that these short NPs can be comfortably accommodated in the pockets between two CSs in eukaryotes and archaea, reducing the internal energy of the NP and slowing the escape to the point of capture. 

In bacteria, on the other hand, the absence of the second constriction in the tunnel enhances the folding and knotting of NPs compared to the other ribosomes \cite{Dabrowski-Tumanski18}. In MD simulations of a cylindrical exit tunnel model, a possible mechanism was described that aids the knotting of deeply knotted NPs during the escape process \cite{Dabrowski-Tumanski18}. In this way, the ribosome interacts with the N-terminal section outside the exit tunnel, holding a twisted loop structure, while the C-terminal section is pulled out and threaded through this loop, forming a knotted protein.

\section*{Concluding Remarks}

Protein synthesis is a ubiquitous, iterative, multistep process in all known life. Here, we dissect the birth of a protein into three stages, which we define on the basis of the protein-termini context. Initially, the synthesized protein passes through the ribosome exit tunnel in an extended conformation, typically too flexible to be accurately modeled using conventional biophysical techniques such as X-ray crystallography or cryo-EM. This presents a fundamental challenge in studying proteosynthesis, and a combination of multiple biophysical techniques is needed \cite{Bock23}. Subsequently, the nascent polypeptide begins to fold, with the folding process often completing only after the polypeptide is released from the ribosome. Only at this point can the polypeptide, now a functional protein, acquire its biological role.

The transition between Stages I and II remains indistinct because of the varied nature of proteins and subtle differences in their synthesis between different species. Likewise, the conclusion of Stage III is also a little ambiguous. In a crowded cellular environment or when associated with a membrane, the mature protein may linger near the ribosome, interacting with its surface, or undergoing a range of post-translational modifications necessary for full maturation.

Furthermore, the complex interplay of translation factors, the ribosome, and nascent polypeptides \cite{Kramer09} continues to elude complete understanding. The functional state of the ribosome surface may vary during protein synthesis. Potential allosteric pathways for information transfer might involve ribosomal proteins such as uL22 \cite{McGrath22} and uL23 \cite{Bornemann08}, although the network of interactions among distant ribosomal components is likely more intricate \cite{Poirot16,Guzel20}. This complexity underscores the need for further research to decipher the nuances of protein synthesis, enhancing our understanding of this fundamental biological process.

\section*{Funding Information}

This work was supported by the Czech Science Foundation (project 23-05557S), by the Ministry of Education, Youth and Sports of the Czech Republic through the e-INFRA CZ (ID: 90254), and by the grants of Specific University Research (A2\_FCHI\_2021\_034,  A1\_FCHI\_2024\_001).

%\bibliographystyle{apalike}
%\bibliography{refs2.bib}
\input{main.bbl}

\end{document}

%% file: main.bbl
\providecommand{\latin}[1]{#1}
\makeatletter
\providecommand{\doi}
  {\begingroup\let\do\@makeother\dospecials
  \catcode`\{=1 \catcode`\}=2 \doi@aux}
\providecommand{\doi@aux}[1]{\endgroup\texttt{#1}}
\makeatother
\providecommand*\mcitethebibliography{\thebibliography}
\csname @ifundefined\endcsname{endmcitethebibliography}
  {\let\endmcitethebibliography\endthebibliography}{}